\documentclass[prd,preprint,
superscriptaddress,showpacs,nofootinbib,%
tightenlines,
]{revtex4}
\usepackage{epsfig}
\usepackage{slashed}
\usepackage{amssymb}

\newcommand{\ben}{\begin{displaymath}}
\newcommand{\een}{\end{displaymath}}
\newcommand{\be}{\begin{equation}}
\newcommand{\ee}{\end{equation}}
\newcommand{\bea}{\begin{eqnarray}}
\newcommand{\eea}{\end{eqnarray}}
\begin{document}
\title{Complex-mass scheme and perturbative unitarity}
    \author{T.~Bauer}
    \affiliation{Institut f\"ur Kernphysik, Johannes
    Gutenberg-Universit\"at, D-55099 Mainz, Germany}
    \author{J.~Gegelia}
    \affiliation{Institut f\"ur Kernphysik, Johannes
    Gutenberg-Universit\"at, D-55099 Mainz, Germany}
    \affiliation{Institut f\"ur Theoretische Physik II, Ruhr-Universit\"at Bochum,
    D-44780 Bochum, Germany}
     \affiliation{High
    Energy Physics Institute, Tbilisi State University, 0186 Tbilisi,
    Georgia}
    \author{G.~Japaridze}
    \affiliation{Clark Atlanta University, Atlanta, GA 30314, USA}
    \author{S.~Scherer}
    \affiliation{Institut f\"ur Kernphysik, Johannes
    Gutenberg-Universit\"at, D-55099 Mainz, Germany}
\date{October 9, 2012}
\begin{abstract}
{ We derive cutting rules for loop integrals containing propagators with complex masses.
   Using a field-theoretical model of a heavy vector boson interacting with a
light fermion, we demonstrate that the complex-mass scheme respects unitarity order by
order in a perturbative expansion provided that the renormalized coupling constant remains real.}

\end{abstract}



\maketitle

\section{Introduction}

   When describing resonances in perturbation theory, one needs to take their finite
widths into account.
   { A framework of handling this problem is provided by the complex-mass scheme
(CMS) \cite{Stuart:1990,Denner:1999gp,Denner:2006ic,Denner:2005fg}.
   This approach has proven successful in various applications within the standard model
\cite{Denner:1999gp,Denner:2006ic,Denner:2005fg,Denner:2006ak,Bredenstein:2006nk,
Bredenstein:2006ha,Actis:2006rc,Actis:2008uh,Actis:2008ug,Denner:2009gj,Denner:2010ia}.
    Recently, the CMS has also been applied in chiral effective field theory
\cite{Djukanovic:2009zn,Djukanovic:2009gt,Bauer:2012at}.

    In the framework of quantum field theory for unstable particles the usage of the CMS
leads to complex-valued renormalized parameters.
   While the problems of unitarity and causality in field theories containing unstable
particles were resolved long ago \cite{Veltman:1963th}, the issue of perturbative unitarity of the $S$-matrix in
the CMS is still open \cite{Denner:2006ic}.
   Using the CMS, one does not change the bare Lagrangian and, therefore, unitarity is not violated in
the complete theory.
   On the other hand, perturbation theory is based on an order-by-order approximation to exact results.
   Therefore, it is not obvious that the approximate expressions for the $S$-matrix also satisfy the
unitarity condition.
    In the present work we examine perturbative unitarity in a model of a heavy abelian vector field
interacting with a light fermion.
   Keeping the renormalized coupling constant (the expansion parameter of the perturbation theory) as a
real quantity, we derive the cutting rules for one-loop integrals involving propagators with complex masses and
show that unitarity is satisfied up to higher-order corrections in the coupling constant.
   In full agreement with Ref.~\cite{Veltman:1963th}, the $S$-matrix connecting only stable states
satisfies the unitarity condition.

\section{Imaginary parts of loop integrals}

   In this section we propose a method for deriving the imaginary parts of (one-loop) integrals
involving propagators with complex masses.
   In the next section we use these results to demonstrate the perturbative unitarity
at one-loop order in a model of a heavy vector boson interacting with a
light fermion.

   Before discussing the propagation of unstable particles, let us recall a few
properties of the Feynman propagator of a stable particle.
   In momentum space the
Feynman propagator of a scalar particle with mass $m$ and four-momentum $p$ is given by\footnote{Recall
the representation of the Dirac delta function,
\begin{displaymath}
\pi\delta(x)=\frac{\epsilon}{x^2+\epsilon^2}\,,\quad\epsilon\to 0^+\,.
\end{displaymath}
}
\begin{eqnarray}
\Delta_F(p) & = & \frac{1}{p^2-m^2+i\,\epsilon} = \frac{p^2-m^2-i\,\epsilon}{\left(p^2-m^2\right)^2+\epsilon^2}
=\frac{p^2-m^2}{\left(p^2-m^2\right)^2+\epsilon^2}-\frac{i\,\epsilon}{\left(p^2-m^2\right)^2+\epsilon^2}\nonumber\\
&=&  \frac{p^2-m^2}{\left(p^2-m^2\right)^2+\epsilon^2}-i\,\pi\,\delta\left(p^2-m^2\right)\,.
\label{stprscalar1}
\end{eqnarray}
   In Eq.~(\ref{stprscalar1}), $\epsilon$ is a positively defined quantity and the
limit $\epsilon\to 0^+$ is assumed.
   Defining $E_p=\sqrt{\vec p\,^2+ m^2}$,
the Feynman propagator has two simple poles at $p_0=E_p-i\epsilon$ and $p_0=-E_p+i\epsilon$, respectively:
\begin{equation}
\Delta_F(p)=\frac{1}{2\,E_p}\left(\frac{1}{p_0-E_p+i\epsilon}-\frac{1}{p_0+E_p-i\epsilon}\right).
\end{equation}
   Besides the Feynman propagator, let us introduce the advanced and retarded propagators,
respectively,
\begin{eqnarray}
\Delta_A(p) & = & \frac{1}{2 \,E_p} \left( \frac{1}{p_0-E_p-i\,\epsilon}-
\frac{1}{p_0+E_p-i\,\epsilon}\right)\nonumber\\
&=& \frac{p^2-m^2}{\left(p^2-m^2\right)^2+\epsilon^2}+i\,\pi\,\delta\left(p^2-m^2\right)\sigma_p,
\label{SAst}\\
\Delta_R(p) & = & \frac{1}{2 \,E_p} \left( \frac{1}{p_0-E_p + i\,\epsilon}-
\frac{1}{p_0+E_p + i\,\epsilon}\right)\nonumber\\
&=&  \frac{p^2-m^2}{\left(p^2-m^2\right)^2+\epsilon^2} - i\,\pi\,\delta\left(p^2-m^2\right)\sigma_p,
\label{SRst}
\end{eqnarray}
where  $\sigma_p\equiv{\rm sign} (p_0)=-{\rm sign} (-p_0)=-\sigma_{-p}$.
   Note that $\Delta_A(p)=\Delta_R(-p)$.
   In terms of these propagators, the Feynman propagator can be written as
\begin{equation}
\Delta_F(p)=\Theta(p_0)\Delta_R(p)+\Theta(-p_0)\Delta_A(p),
\end{equation}
with the Heaviside step function $\Theta(t)=1$ for $t>0$ and $\Theta(t)=0$ for $t<0$.

{    In the CMS, for an unstable scalar particle with mass $M$, width $\Gamma$, and
four-momentum $p$, the Feynman propagator of Eq.~(\ref{stprscalar1}) is replaced by}\footnote{
We stress that in our notation the prime does not refer to the full, dressed propagator of the
unstable particle.}
\begin{eqnarray}
\Delta'(p) & = & \frac{1}{p^2-M^2+i\,(M\Gamma+\epsilon)}
\nonumber\\
& = & \frac{p^2-M^2}{\left(p^2-M^2\right)^2+(M\,\Gamma+\epsilon)^2}-\frac{i\,(M\Gamma+\epsilon)}
{\left(p^2-M^2\right)^2+(M\Gamma+\epsilon)^2}\,.
\label{CMSprscalar}
\end{eqnarray}
    For a finite width $\Gamma$, the infinitesimal parameter $\epsilon$ can be neglected.
    If one is also interested in the case of vanishing $\Gamma$, i.e.~stable particles, one has to keep
the infinitesimal parameter $\epsilon$.
   In the following, we drop $\epsilon$ with the understanding that it is easily reintroduced
by replacing $M\,\Gamma\to M\,\Gamma+\epsilon$.
    Let us define auxiliary functions which we denote as ''advanced'' and ''retarded'' propagators
$\Delta'_A$ and $\Delta'_R$ corresponding to
$\Delta'$:
\begin{eqnarray}
\Delta'_A(p) & = & \frac{1}{w(p)+w(p)^*} \left[ \frac{1}{p_0-w(p)^*}- \frac{1}{p_0+w(p)}\right]\nonumber\\
& = & \frac{p^2-M^2-M^2 \Gamma^2/(2\,x(p)^2)}{\left(p^2-M^2\right)^2+M^2\Gamma^2}+
\frac{i\,M\,\Gamma}{\left(p^2-M^2\right)^2+M^2\Gamma^2}\,\frac{p_0}{x(p)}\,,
\label{SAust}\\
\Delta'_R(p) & = & \frac{1}{w(p)+w(p)^*} \left[ \frac{1}{p_0-w(p)}-
\frac{1}{p_0+w(p)^*}\right]\nonumber\\
& = &  \frac{p^2-M^2-M^2 \Gamma^2/(2\,x(p)^2)}{\left(p^2-M^2\right)^2+M^2\Gamma^2}
- \frac{i\,M\,\Gamma}{\left(p^2-M^2\right)^2+M^2\Gamma^2}\,\frac{p_0}{x(p)}\,,
\label{SRust}
\end{eqnarray}
where
\begin{eqnarray}
w(p) & = & x(p)-i\,y(p)\,,\nonumber\\
x(p) & = & \frac{1}{\sqrt{2}}\,\sqrt{\left({\cal E}_p^4+M^2\Gamma^2\right)^{1/2}+{\cal E}_p^2}
={\cal E}_p+{\cal O}(\Gamma^2)\,,\nonumber\\
y(p) & = & \frac{1}{\sqrt{2}}\,\sqrt{\left({\cal E}_p^4+M^2\Gamma^2\right)^{1/2}-{\cal E}_p^2}
= M\,\Gamma/(2\,{\cal E}_p)+{\cal O}(\Gamma^3)\,,\nonumber\\
{\cal E}_p & = & \sqrt{\vec{p}\,^2 + M^2}\,.
\label{z1def}
\end{eqnarray}
{  Here and below it is understood that $\Gamma\ll M$ and hence the small expansion parameter is $\Gamma/M$.}
   The ''advanced'' (''retarded'') propagator has two simple poles in the upper (lower) complex
half plane and approaches the advanced (retarded) propagator of a stable particle as
$\Gamma\to 0$.

   With the above definitions, let us consider a generic one-loop integral involving the
propagation of both a stable and an unstable particle,
\begin{eqnarray}
I_1 = i \int \frac{d^4 k}{(2 \pi)^4}\frac{d^4 q}{(2 \pi)^4}\,(2 \pi)^4 \delta^4(k+q-p) \Delta_F(k) \Delta'(q).
\label{LoopI}
\end{eqnarray}
   Using Eqs.~(\ref{stprscalar1}) and (\ref{CMSprscalar}), the imaginary part of the integral $I_1$ is
given by
\begin{eqnarray}
{\rm Im } [I_1] & = & \int \frac{d^4 k}{(2 \pi)^4}\frac{d^4 q}{(2 \pi)^4}\,(2 \pi)^4 \delta^4(k+q-p)
\Biggl[\frac{k^2-m^2}{\left(k^2-m^2\right)^2+\epsilon^2}\,\frac{q^2-M^2}{\left(q^2-M^2\right)^2+M^2\Gamma^2} \nonumber\\
&&-\pi\,\delta\left(k^2-m^2\right) \frac{M\,\Gamma}{\left(q^2-M^2\right)^2+M^2\Gamma^2}  \Biggr]. \label{LoopIIm}
\end{eqnarray}
   The purpose of the subsequent manipulations is to bring the first term on the right-hand side of
Eq.~(\ref{LoopIIm}) into a more convenient form.
   To that end, we make use of the following observation.
   As functions of the complex variable $q_0$, both $\Delta_R(q-p)$ and $\Delta'_R(q)$ have simple
poles in the lower half plane which is also true for their product.
   Closing the contour integration in the upper half plane including a vanishing contribution
resulting from the semi circle at infinity,
we find, using Cauchy's theorem,
\begin{eqnarray}
0&=&i\int\frac{d^4 q}{(2 \pi)^4}\,\Delta_R(q-p) \Delta_R'(q)\nonumber\\
&=&i\int\frac{d^4 q}{(2 \pi)^4}\,\Delta_A(p-q) \Delta_R'(q)\nonumber\\
&=&i \int \frac{d^4 k}{(2 \pi)^4}\frac{d^4 q}{(2 \pi)^4}\,(2 \pi)^4 \delta^4(k+q-p) \Delta_A(k) \Delta_R'(q)\,.
\label{LoopI0}
\end{eqnarray}
   By substituting Eqs.~(\ref{SAst}) and (\ref{SRust}) into Eq.~(\ref{LoopI0}) and taking the imaginary parts
of both sides, we obtain
\begin{eqnarray}
0  & = & \int \frac{d^4 k}{(2 \pi)^4}\frac{d^4 q}{(2 \pi)^4}\,(2 \pi)^4 \delta^4(k+q-p)
\Biggl[\frac{k^2-m^2}{\left(k^2-m^2\right)^2+\epsilon^2}\,\frac{q^2-M^2-M^2 \Gamma^2 /(2 \,x(q)^2)}{\left(q^2-M^2\right)^2
+M^2\Gamma^2} \nonumber\\ &&+
\pi\,\delta\left(k^2-m^2\right)\,\sigma_k\,
\frac{M\,\Gamma\,}{\left(q^2-M^2\right)^2+M^2\Gamma^2} \frac{q_0}{x(q)}\, \Biggr].
\label{0imp}
\end{eqnarray}
   Expanding $x(q)={\cal E}_q+{\cal O}(\Gamma^2)$, we can replace Eq.~(\ref{0imp}) by
\begin{eqnarray}
0 & = & \int \frac{d^4 k}{(2 \pi)^4}\frac{d^4 q}{(2 \pi)^4}\,(2 \pi)^4 \delta^4(k+q-p)
\Biggl[\frac{k^2-m^2}{\left(k^2-m^2\right)^2+\epsilon^2}\,\frac{q^2-M^2}{\left(q^2-M^2\right)^2+M^2\Gamma^2} \nonumber\\
&&+\pi\,\delta\left(k^2-m^2\right)\,\sigma_k\, \frac{M\,\Gamma\,
}{\left(q^2-M^2\right)^2+M^2\Gamma^2}\frac{q_0}{{\cal E}_q}+{\cal O}(\Gamma^2)  \Biggr]\,.
\label{LoopIImzero}
\end{eqnarray}
   Subtracting Eq.~(\ref{LoopIImzero}) from Eq.~(\ref{LoopIIm}) and restoring $\epsilon$,
we obtain
\begin{equation}
{\rm Im } [I_1] = -\pi \int \frac{d^4 k}{(2 \pi)^4}\,\left[
\delta\left(k^2-m^2\right)
\frac{
(M\,\Gamma+\epsilon)\,
\left(1
+{\sigma_k}\,\frac{p_0-k_0}{{\cal E}_{p-k}}
\right)
}{\left[(p-k)^2-M^2\right]^2+(M\,\Gamma+\epsilon)^2}
+{\cal O}(\Gamma^2)\right].
\label{a}
\end{equation}
  It is convenient to rewrite Eq.~(\ref{a}) in a form analogous to the
standard cutting formula for loop integrals with real masses \cite{Cutkosky:1960sp,Peskin:1995ev}.
   The left-hand side of Eq.~(\ref{a}) is manifestly Lorentz invariant, and hence the right-hand
side is Lorentz invariant order by order in $\Gamma$.
   Based on this fact we derive that
\begin{equation}
{\rm Im } [I_1]
=  -\pi \int \frac{d^4 k}{(2 \pi)^4}\,
\delta\left(k^2-m^2\right)
\frac{
(M\,\Gamma+\epsilon)\,
\left(1+{\sigma_k}\,\sigma_{p-k}\right)
}{\left[(p-k)^2-M^2\right]^2+(M\,\Gamma+\epsilon)^2}
+{\cal O}(\Gamma^2).
\label{LoopIImFinal}
\end{equation}
{    The symbol ${\cal O}(\Gamma^2)$ on the right-hand side of Eq.~(\ref{LoopIImFinal}) indicates that
the neglected terms contain at least one additional overall factor of $\Gamma$ as compared to the first term.
   Note that the integrand of Eq.~(\ref{LoopIImFinal}) is not obtained by expanding the integrand
of Eq.~(\ref{a}).
   The equivalence of expressions (\ref{a}) and (\ref{LoopIImFinal}) can rather be seen
by subtracting them from each other and considering, for $\epsilon=0$, the limit $\Gamma\to 0$ of
the integrated result.
   Both Eq.~(\ref{LoopIImFinal}) and Eq.~(\ref{a}) turn into the expression for the stable
particles in this limit.
   Hence the difference has to be suppressed by an additional factor of $\Gamma$.}

   The generalization of the above procedure to any one-loop integral
containing propagators with complex masses is straightforward.
   In analogy to Eq.~(\ref{LoopIImFinal}), cutting an unstable-particle line
results in an overall factor of $\Gamma$, whereas cutting a stable-particle line
generates a delta function.
   For integrals containing propagators of stable particles only,
the usual cutting rules apply.
   A list of the imaginary parts of one-loop integrals needed for
the calculations of the next sections is given in the appendix.

\section{The Model}

    In the next section we demonstrate that using complex renormalized masses in unstable-particle
propagators does not violate perturbative unitarity.
   To that end let us consider a model describing the interaction of an unstable vector boson
($B$) with a stable fermion ($\psi$),
\begin{eqnarray}
{\cal L} & = & -{1\over 4} \ F_{0 \mu\nu} F_0^{\mu\nu}
+\frac{M_0^2}{2}\, B_{0 \mu} B_0^{\mu}
+ \bar \psi_0 \left(i\slashed{\partial}-m_0\right) \psi_0
+ g_0\, \bar \psi_0 \,\gamma_\mu \psi_0\, B_0^\mu , \label{EffLagr}
\end{eqnarray}
where $F_{0\mu\nu}=\partial_\mu B_{0\nu} - \partial_\nu B_{0\mu}$.
   The subscript 0 indicates bare parameters and fields.
   The masses are chosen such that the vector boson can decay into a fermion-antifermion pair.
   Since the vector boson couples to a conserved vector current, the above model is
on-mass-shell renormalizable \cite{Boulware:1970zc}, i.e.~leads to finite physical
quantities after renormalizing the masses and the coupling constant.

   Our perturbative approach to the considered model is based on the path integral formalism.
   The integration over classical fields corresponding to particles (stable as well as unstable)
is performed in the standard way, i.e., the Gaussian part is treated non-perturbatively and the rest
perturbatively. { For stable particles, perturbation theory based on the path integral formalism is equivalent
to perturbation theory based on the operator formalism in the Dirac interaction representation.}
   On the other hand, the functional integral also allows to incorporate the unstable degrees of freedom,
while the application of the interaction picture using a ``free'' Hamiltonian for unstable states
is conceptually problematic.

    We perform the renormalization in two steps: first we get rid
off the ultraviolet divergences by applying dimensional regularization in
combination with the $\overline{\rm MS}$ scheme \cite{Collins:1984xc}.
   We refrain from showing the corresponding counter terms explicitly
(including those leading to the wave function renormalization).
   Next we express the renormalized masses of the $\overline{\rm MS}$ scheme in terms
of physical quantities---the poles of the dressed propagators---and substitute them
back into the Lagrangian.
   This amounts to performing the following substitutions in Eq.~(\ref{EffLagr}),
\begin{eqnarray}
B_0^\mu & \to & B^\mu,\nonumber\\
\psi_0 & \to & \psi,\nonumber\\
m_0 & \to & m+\delta m,\nonumber\\
 M_0^2 & \to & M^2-i\,M \Gamma +\delta z \equiv z+\delta z \,,\nonumber\\
 g_0 & \to & g,
\label{renpar}
\end{eqnarray}
resulting in
\begin{eqnarray}
{\cal L} & = & {\cal L}_{\rm main} +{\cal L}_{\rm ct}\,,\nonumber\\
{\cal L}_{\rm main} & = & -{1\over 4} \ F_{\mu\nu} F^{\mu\nu}
+\frac{z}{2}\, B_{\mu} B^{\mu}+ \bar \psi \left( i
\slashed{\partial} -m\right)
\psi + g\, \bar \psi \,\gamma_\mu \psi\, B^\mu ,\nonumber\\
{\cal L}_{\rm ct} & = & \frac{\delta z}{2}\, B_{\mu} B^{\mu}- \delta m\, \bar \psi\,
\psi\,.
\label{EffLagrRewritten}
\end{eqnarray}
    The main Lagrangian ${\cal L}_{\rm main}$ generates the propagators and the vector-boson fermion
interaction vertex.
   The counter-term Lagrangian ${\cal L}_{\rm ct}$ is treated perturbatively in a loop expansion,
i.e., we write the counter terms as
$\delta m=\sum_{k=1}^\infty \hbar^k \delta m_k$ and $\delta z=\sum_{k=1}^\infty \hbar^k \delta z_k$
and include them order by order in perturbative calculations.

    The undressed propagators of the fermion and the vector boson take the following form, respectively,
\begin{eqnarray}
S_F(p) & = & \frac{1}{\slashed{p}-m+i\,\epsilon}\,,\label{Fprop}\\
S_{\mu\nu}'(p) & = & -\,\frac{g_{\mu\nu}- p_\mu p_\nu/z}{p^2 - z},
\label{Vprop}
\end{eqnarray}
and, for later usage, we parameterize the self energy of the vector boson as
\begin{equation}
i\,\Pi_{\mu\nu}(p) = i \left[\Pi(p^2)g_{\mu\nu}+\Pi_p(p^2)p_\mu p_\nu\right ].
\label{vse}
\end{equation}

\section{Perturbative unitarity of the $S$-matrix}

     Below we demonstrate that the unitarity condition for
the forward-scattering amplitude,
\begin{equation}
\label{unitaritycondition}
{\rm Im}\left[{\cal T}_{ii}\right]=\frac{1}{2}\sum_n\hspace{-1.4em}\int\hspace{0.5em}
(2\pi)^4\delta^4(P_n-P_i){\cal T}_{ni}^\ast{\cal T}_{ni},
\end{equation}
is satisfied at one-loop order
in perturbation theory.
   In Eq.~(\ref{unitaritycondition}), the $T$-matrix element between
initial and final four-momentum eigenstates is written as
$\langle f|T|i\rangle=(2\pi)^4\delta^4(P_f-P_i){\cal T}_{fi}$.
   The imaginary parts of one-loop integrals needed in this section are given in the appendix.

   We start with the tree-order amplitude of the process $f(p_1,\sigma_1)+\bar f(p_2,\sigma_2)\to
f(p_3,\sigma_3)+\bar f(p_4,\sigma_4)$ shown in Fig.~1 a),
\begin{eqnarray}
i\,{\cal T}_a & = & -i\,g^2\,\bar u(p_3,\sigma_3)\gamma^\mu
v(p_4,\sigma_4)\, S'_{\mu\nu}(p)\, \bar v(p_2,\sigma_2)\gamma^\nu u(p_1,\sigma_1)
\equiv - i\,  V^\mu_f \,
S'_{\mu\nu}(p)\,V^\nu_i\,,\label{ampltree}
\end{eqnarray}
where $p=p_1+p_2=p_3+p_4$ with $p^2=s$ and our fermion states are
normalized as
$\langle f(\vec{p}\,',\sigma')|f(\vec{p},\sigma)\rangle
=2 E(\vec p)(2\pi)^3\delta(\vec p\,'-\vec p)\delta_{\sigma'\sigma}$
(and analogously for antifermion states);
   the Dirac spinors are normalized as $\bar{u}(\vec p,\sigma)u(\vec p,\sigma)
=2m=-\bar{v}(\vec p,\sigma)v(\vec p,\sigma)$.
   Because of current conservation, using Eq.~(\ref{Vprop}), we obtain
from Eq.~(\ref{ampltree})
\begin{equation}
{\cal T}_a  =   V^\mu_f \,\frac{1}{s-z}\,V_{i\mu}\,.\label{ampltreeequiv}
\end{equation}
   From now on we consider forward scattering, i.e., $p_1=p_3$, $p_2=p_4$, $\sigma_1=\sigma_3$,
and $\sigma_2=\sigma_4$.
   Renaming $V_{i\mu}\to V_\mu$, and using $V^\mu_f=V^{\mu\ast}$,
the imaginary part of the forward-scattering tree-order
amplitude obtained from Eq.~(\ref{ampltreeequiv}) reads
\begin{equation}
{\rm Im} \left[{\cal {T}}_a\right]  = {\rm Im} \left[V^{\mu\ast} \frac{1}{s-z}V_\mu\right]
={\rm Im} \left[ V^{\mu\ast} \,\frac{s-z^*}{(s-z)(s-z^*)}\,V_\mu \right] =- V^{\mu\ast}
\,\frac{M\,\Gamma}{(s-z)(s-z^*)}\,V_\mu\,.\label{ImPartampltree}
\end{equation}
  We omit wave function renormalization constants for external fermion
lines as they do not contribute to the obtained relations at the given accuracy.

    The one-loop diagrams contributing to the $f\bar f \to f\bar f$ amplitude are
shown in Fig.~\ref{oneloop:fig}.
\begin{figure}
\epsfig{file=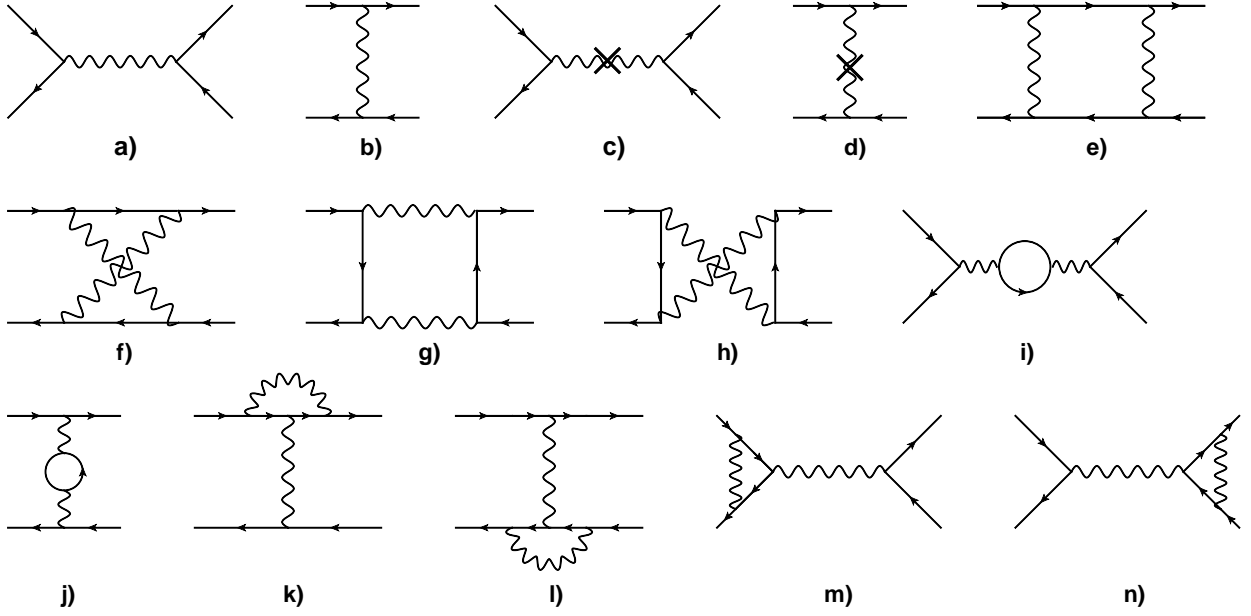, width=\textwidth}
\caption[]{\label{oneloop:fig} Tree and
one-loop contributions to $f\bar f$ scattering. Solid and curved lines
correspond to fermions and vector bosons, respectively.
   Crosses refer to counter-term contributions.}
\end{figure}
   Let us begin with discussing the imaginary parts of
diagrams c) and i) relevant for the calculation up to and including order $\Gamma^2$.
   Throughout we exploit the fact that $g\sim \sqrt{\Gamma}$ (hence also $V_\mu\sim\sqrt{\Gamma}$).
   The results of these two diagrams read
\begin{eqnarray}
i\,{\cal T}_{i} & = &
i\,V^{\mu\ast} \,  S'_{\mu\alpha}(p)\,
\Pi^{\alpha\beta}_1(p)\, S'_{\beta\nu}(p)\,V^\nu ,\nonumber\\
i\,{\cal T}_c & = & i\,V^{\mu\ast} \,  S'_{\mu\alpha}(p)\, \delta z_1
\,g^{\alpha\beta}\,
S'_{\beta\nu}(p)\,V^\nu\,,\label{amplloop}
\end{eqnarray}
  where $\Pi^{\alpha\beta}_1(p)$ and $\delta z_1$ denote the self energy and
the counter term at first order in $\hbar$, respectively.
   Using Eq.~(\ref{vse}), we obtain from Eq.~(\ref{amplloop})
\begin{equation}
{\cal T}_{i+c}  = {\cal T}_i+{\cal T}_c =  V^{\mu\ast} \,\frac{\Pi_1(s)+\delta z_1}{(s-z)^2}\,V_\mu\,,
\label{amplloopequiv}
\end{equation}
with $\Pi_1(s)$ denoting the one-loop contribution to the function $\Pi(s)$ of Eq.~(\ref{vse}).
   The imaginary part of the one-loop order amplitude for $s\neq M^2$ is obtained from Eq.~(\ref{amplloopequiv})
as
\begin{equation}
{\rm Im} \left[{\cal T}_{i+c}\right]  =  V^{\mu\ast} \,\frac{{\rm Im} \left[
(\Pi_1(s)+\delta z_1) (s-z^*)^2\right]}{(s-z)^2(s-z^*)^2}\,V_\mu =
V^{\mu\ast} \,\frac{{\rm Im} \left[ \Pi_1(s)+\delta
z_1\right]}{(s-z)(s-z^*)}\,V_\mu +{\cal
O}\left(\Gamma^3\right)\,.\label{ImPartamplloop}
\end{equation}
 In Eq.~(\ref{ImPartamplloop}) we took into account that $g\sim \sqrt{\Gamma}$ and hence $V_\mu\sim\sqrt{\Gamma}$,
$\Pi_1(s)\sim\Gamma$, and $\delta z_1\sim\Gamma$.

   Let us now consider the contribution of the intermediate state consisting
of one fermion ($f$) and one antifermion ($\bar{f}$) ( ''square'' of the tree-order amplitude shown
in Fig.~\ref{mnloops:fig}) to the right-hand side of the unitarity
condition of Eq.~(\ref{unitaritycondition}):
\begin{eqnarray}
\lefteqn{\frac{1}{2}\sum_{f,\bar{f}}\hspace{-1.4em}\int\hspace{0.5em}
(2\pi)^4\delta^4(p_{(f\bar{f})}-p){\cal T}_{(f\bar{f})i}^\ast{\cal T}_{(f\bar{f})i}}\nonumber\\
&= &\frac{1}{2}g^4\,\sum_{\sigma,\bar{\sigma}}\int \frac{d^4 q}{(2\pi)^4} \,
(2\pi)\delta\left[q^2-m^2\right]\Theta(q_0)\,
(2\pi)\delta\left[(p-q)^2-m^2\right]\Theta(p_0-q_0)\nonumber\\
&& \times\bar u(p_1,\sigma_1)\gamma_\nu v(p_2,\sigma_2)
\left[S^{'\beta\nu}(p)\right]^\ast\bar v(p-q,\bar\sigma)\gamma_\beta u(q,\sigma)\nonumber\\
&&\times \bar u(q,\sigma)\gamma_\alpha v(p-q,\bar\sigma)\,
S^{'\alpha\mu}(p) \, \bar v(p_2,\sigma_2)\gamma_\mu
u(p_1,\sigma_1) \nonumber\\
&=&2 \pi^2 g^2\,\int \frac{d^4 q}{(2\pi)^4} \,
\delta\left[q^2-m^2\right]\Theta(q_0)\,
\delta\left[(p-q)^2-m^2\right]\Theta(p_0-q_0)\nonumber\\
&&\times\frac{1}{s-z^\ast}V_\nu^\ast\,\mbox{Tr}\left[\gamma^\nu(\slashed{q}+m)\gamma^\mu
(\slashed{p}-\slashed{q}-m)\right]
\frac{1}{s-z}V_\mu\nonumber\\
&=& V^{\ast\mu} \,\frac{{\rm Im}\left[ \Pi_1(s)\right]}{(s-z)(s-z^*)}\,V_\mu\,.
\label{ttdaggerTree}
\end{eqnarray}
\begin{figure}
\epsfig{file=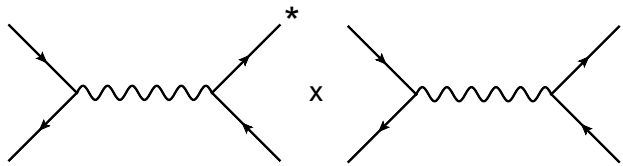, width=0.5\textwidth}
\caption[]{\label{mnloops:fig} Squared s-channel tree-order $f \bar f$ scattering
amplitude. Solid and curved lines correspond to
fermions and vector bosons, respectively, and the star indicates the complex conjugate of
the amplitude represented by the corresponding diagram.}
\end{figure}
   In the second-to-last step we made use of the projection operators over the positive
and negative energy states.
   The last step is obtained by applying the cutting rules 
for stable particles to the one-loop self energy of the vector boson,
\begin{equation}
\label{Pioneloop}
\Pi^{\mu\nu}_1(p)=i g^2 \int\frac{d^4 q}{(2\pi)^4}\mbox{Tr}
\left(\frac{1}{\slashed{q}-\slashed{p}-m+i\epsilon}\gamma^\mu\frac{1}{\slashed{q}-m+i\epsilon}\gamma^\nu
\right).
\end{equation}
   Using $1+\sigma_q\sigma_{p-q}=2[\Theta(q_0)\Theta(p_0-q_0)+\Theta(-q_0)\Theta(q_0-p_0)]$,
the fact that $p_0\geq 2m$ in Eq.~(\ref{ttdaggerTree}), and Eq.~(\ref{vse}), results
in Eq.~(\ref{ttdaggerTree}).
   Let us compare Eq.~(\ref{ttdaggerTree}) with the sum of
Eqs.~(\ref{ImPartampltree}) and (\ref{ImPartamplloop}).
   Taking into account that ${\rm Im}[\delta z_1]= M\,\Gamma +{\cal O}(\hbar^2)$
[which follows from Eq.~(\ref{renpar})], we see that the unitarity condition
is satisfied up to and including order $\Gamma^2$.

  In the following, we qualitatively discuss the imaginary parts of the remaining diagrams of
Fig.~\ref{oneloop:fig} using the formulae in the appendix.
   The sum of diagrams b) and d) may be expanded in powers of $\Gamma$.
   The first term in this expansion is real, whereas at second order the respective
imaginary parts of the two diagrams cancel each other.
   Therefore, the corresponding imaginary part of these two diagrams is of ${\cal O}(\Gamma^3)$.
   Similarly, expanding the amplitude corresponding to diagram j), the first term in this expansion
is of order $g^4$ and is real and the imaginary part starts contributing at ${\cal O}(\Gamma^3)$.
   The imaginary parts of diagrams f), g), h), k), and l) are all of higher order.
   This is the case because all these diagrams are proportional to $g^4$ and applying vertical
cuts (relevant to forward scattering) in each case means cutting at least one unstable-particle line
producing an additional factor of $\Gamma$.
   Finally, cutting two stable fermion lines in diagrams e), m), and n) generates the imaginary parts
corresponding, up to higher-order terms, to the ''square'' of the tree-order diagrams
shown in Fig.~\ref{mnloops2:fig}.
   On the other hand, all remaining possible cuts involve at least one unstable line which leads
at least to ${\cal O}(\Gamma^3)$.

\begin{figure}
\epsfig{file=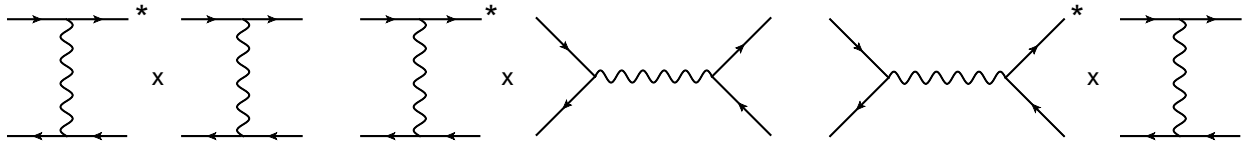,width=\textwidth}
\caption[]{\label{mnloops2:fig} Squared $t$-channel, $s$-channel $t$-channel
interference,  $t$-channel $s$-channel interference tree-order
amplitudes. Solid and curved lines correspond to
fermions and vector bosons, respectively, and the star indicates the complex conjugate of
the amplitude represented by the corresponding diagram.}
\end{figure}

    According to Ref.~\cite{Veltman:1963th}, only stable asymptotic states contribute to
the unitarity condition of the $S$-matrix.
   In order to demonstrate that our cutting rules agree with this result,
we consider the one-loop contribution to the forward-scattering amplitude
of $ f f \bar f\to  f f \bar f$ shown in Fig.~\ref{VF:fig} a).
   Using Eq.~(\ref{LoopIImFinal}), it is easily seen that the imaginary part obtained by cutting the two
lines of the loop in diagram a) is proportional to $\Gamma$.
   In Fig.~\ref{VF:fig} b), it is schematically represented as the ''square'' of the tree-order diagram
(modulo higher-order corrections).
   As the width $\Gamma$ is generated by diagrams representing the decay of the vector boson
into stable particles, it is clear that the imaginary part of the diagram in Fig.~\ref{VF:fig} a)
corresponds to the ''square'' of diagrams with stable particles only in external legs [see
diagram c) in Fig.~\ref{VF:fig}].
   Since the width $\Gamma$ is calculated for an ''on-mass-shell'' vector boson, diagram c) contains also
contributions corresponding to loop diagrams of higher order.
   Note that in the limit of vanishing $\Gamma$ ($M<2m$) one is not allowed to drop $\epsilon$ in Eq.~(\ref{LoopIImFinal}).
   Therefore, the result of diagram \ref{VF:fig} b) does not vanish.
   In fact, for $\Gamma=0$ the limit $\epsilon\to 0$ leads to a delta function corresponding to the vector
line, and we obtain the standard cutting rule for stable particles.

\begin{figure}
\epsfig{file=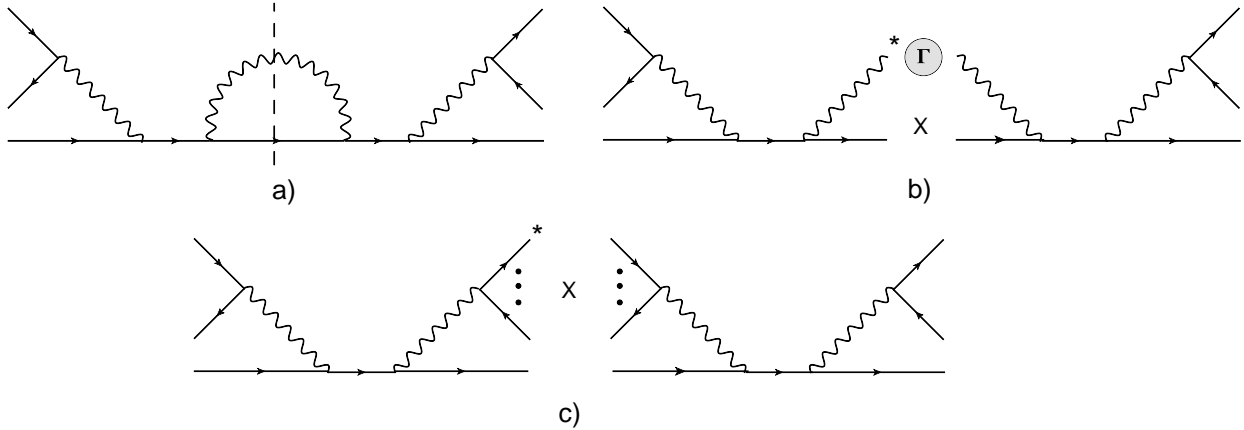, width=\textwidth}
\caption[]{\label{VF:fig} $ f f \bar f$ scattering. a) one-loop
contribution, b) ''square'' of the tree-order diagram, c) diagrams
with stable particles only in external legs. Solid and curved lines
correspond to fermions and vector bosons, respectively, and the star
indicates the complex conjugate of the amplitude represented by the
corresponding diagram.}
\end{figure}

\section{Conclusions}

   In this work we developed a procedure for deriving the imaginary parts of (one-loop) integrals
involving propagators with complex masses.
   With the aid of this method, we demonstrated perturbative unitarity of the
scattering amplitude within the complex-mass scheme at the one-loop level.
   This result was obtained under the assumption that the expansion parameter of
perturbation theory (the renormalized coupling constant) remains real.
   Our results are in full agreement with the findings of Ref.~\cite{Veltman:1963th} that
unstable states do not appear as asymptotic states and are therefore excluded from the unitarity condition.
    A generalization of cutting rules for unstable particles to higher orders of the loop expansion is
straightforward.
   However, because of the non-trivial dependence of imaginary parts on $\Gamma$, the analysis
of perturbative unitarity in higher orders will become more involved.

\acknowledgments

J.~G.~acknowledges the support of
the Deutsche Forschungsgemeinschaft (SFB 443) and Georgian National
Foundation grant GNSF/ST08/4-400.
T.~B.~was supported by the Deutsche Forschungsgemeinschaft (SCHE459/4-1) and the
German Academic Exchange Service (DAAD).
J.~G.~and T.~B.~would like to thank M.~Paris for discussions and comments on the manuscript.
T.~B.~would like to thank H.~W.~Grie{\ss}hammer and M.~R.~Schindler for useful discussions and their
hospitality during his stay at George Washington University.

\appendix

\section{Imaginary parts of one-loop integrals}
In order to compactify the notation let us introduce the following abbreviations,
\begin{eqnarray*}
d(k^2)&=&(k^2-m^2)^2+\epsilon^2,\\
D(k^2)&=&(k^2-M^2)^2+M^2\Gamma^2.
\end{eqnarray*}

We consider the integral
\begin{eqnarray}
I_2 & = & i \int \frac{d^4 k}{(2 \pi)^4}\frac{d^4 q}{(2 \pi)^4}\frac{d^4 l}{(2 \pi)^4}
\,(2 \pi)^4 \delta^4(k+l-p_1)\,(2 \pi)^4 \delta^4(k+q-p_2) \Delta'(k) \Delta_F(q)\Delta_F(l)\nonumber\\
& = & i\int \frac{d^4 k}{(2 \pi)^4}
\,\Delta'(k) \Delta_F(p_2-k)\Delta_F(p_1-k)\,.
\label{LoopI3}
\end{eqnarray}
    Its imaginary part reads
\begin{eqnarray}
{\rm Im }[I_2] & = & \int \frac{d^4 k}{(2 \pi)^4}\frac{d^4 q}{(2
\pi)^4}\frac{d^4 l}{(2 \pi)^4}
\,(2 \pi)^4 \delta^4(k+l-p_1)\,(2 \pi)^4 \delta^4(k+q-p_2)\nonumber\\
&& \times \Biggl\{ \frac{k^2-M^2}{D(k^2)}\,\frac{q^2-m^2}
{d(q^2)}\,\frac{l^2-m^2} {d(l^2)}- \frac{k^2-M^2}{D(k^2)}\,\pi^2
\delta(q^2-m^2)\,\delta(l^2-m^2)\nonumber\\
&& - \frac{M\,\Gamma}{D(k^2)} \,\frac{q^2-m^2}{d(q^2)}\,\pi\,
\delta(l^2-m^2) - \frac{M\,\Gamma}{D(k^2)}\,
\frac{l^2-m^2}{d(l^2)}\,\pi\, \delta(q^2-m^2) \Biggr\}\,.
\label{LoopI3I}
\end{eqnarray}
    To rewrite the expression of Eq.~(\ref{LoopI3I}) in a more convenient form
we consider the following integral,
\begin{eqnarray}
0 = i \int \frac{d^4 k}{(2 \pi)^4}\frac{d^4 q}{(2 \pi)^4}\frac{d^4 l}{(2 \pi)^4}
\,(2 \pi)^4 \delta^4(k+l-p_1)\,(2 \pi)^4 \delta^4(k+q-p_2) \Delta_A'(k) \Delta_R(q)\Delta_R(l)\,,
\label{LoopI3zero}
\end{eqnarray}
take its imaginary part and obtain
\begin{eqnarray}
0 & = & \int \frac{d^4 k}{(2 \pi)^4}\frac{d^4 q}{(2 \pi)^4}\frac{d^4 l}{(2 \pi)^4}
\,(2 \pi)^4 \delta^4(k+l-p_1)\,(2 \pi)^4 \delta^4(k+q-p_2)\nonumber\\
&& \times \Biggl\{ \frac{k^2-M^2}{D(k^2)}\,\frac{q^2-m^2}
{d(q^2)}\,\frac{l^2-m^2} {d(l^2)} - \frac{k^2-M^2}{D(k^2)}\,\pi^2
\delta(q^2-m^2)\,\delta(l^2-m^2)\,\sigma_q \sigma_l\nonumber\\
&& +  \frac{M\,\Gamma}{D(k^2)} \,\frac{q^2-m^2}{d(q^2)}\,\pi\,
\delta(l^2-m^2)\,\sigma_l\,\frac{k_0}{{\cal E}_k}+
\frac{M\,\Gamma}{D(k^2)}\, \frac{l^2-m^2}{d(l^2)}\,\pi\,
\delta(q^2-m^2)\,\sigma_q\,\frac{k_0}{{\cal E}_k} \nonumber \\
&&+\,{\cal O}(\Gamma^2) \Biggr\}\,. \label{LoopI3Izero}
\end{eqnarray}
   By subtracting Eq.~(\ref{LoopI3Izero}) from Eq.~(\ref{LoopI3I}), we obtain
\begin{eqnarray}
{\rm Im }[I_2] & = & \int \frac{d^4 k}{(2 \pi)^4}\frac{d^4 q}{(2 \pi)^4}\frac{d^4 l}{(2 \pi)^4}
\,(2 \pi)^4 \delta^4(k+l-p_1)\,(2 \pi)^4 \delta^4(k+q-p_2)\nonumber\\
&& \times  \Biggl\{ - \frac{k^2-M^2}{D(k^2)}\,\pi^2
\delta(q^2-m^2)\,\delta(l^2-m^2)\,(1-\sigma_q \sigma_l)\nonumber\\
&& -  \frac{M\,\Gamma\,\left(1+\sigma_l\,\frac{k_0}{{\cal
E}_k}\right)}{D(k^2)} \,\frac{q^2-m^2}{d(q^2)}\,\pi\,
\delta(l^2-m^2)\nonumber\\
&& -  \frac{M\,\Gamma\,\left(1+\sigma_q\,\frac{k_0}{{\cal
E}_k}\right)}{D(k^2)}\, \frac{l^2-m^2}{d(l^2)}\,\pi\,
\delta(q^2-m^2) +{\cal O}(\Gamma^2) \Biggr\}\,. \label{LoopI3IFinal}
\end{eqnarray}
   The first term in Eq.~(\ref{LoopI3IFinal}) corresponds to cutting both stable-particle lines
\cite{Cutkosky:1960sp,Peskin:1995ev}.
   The other two terms correspond to cutting one of the two stable-particle lines together with
the unstable-particle line.
   These two terms are proportional to $\Gamma$.

    Next, let us consider an integral
\begin{eqnarray}
I_3 & = & i\int \frac{d^4 k}{(2 \pi)^4}\frac{d^4 q}{(2 \pi)^4}\frac{d^4 l}{(2 \pi)^4}
\frac{d^4 p}{(2 \pi)^4}
\,(2 \pi)^4 \delta^4(k+p_1-p_3-q)\,(2 \pi)^4 \delta^4(k+p_1-l)\nonumber\\
&& \times  (2 \pi)^4 \delta^4(k-p_2+p) \Delta'(k)\Delta'(q)\Delta_F(l)\Delta_F(p)\nonumber\\
& = & i \int \frac{d^4 k}{(2 \pi)^4}
\,\Delta'(k)\Delta'(p_1-p_3+k)\Delta_F(p_1+k)\Delta_F(p_2-k)\,.
\label{LoopI4}
\end{eqnarray}
\noindent
    Its imaginary part reads
\begin{eqnarray}
{\rm Im }[I_3] & = & \int \frac{d^4 k}{(2 \pi)^4}\frac{d^4 q}{(2 \pi)^4}\frac{d^4 l}{(2 \pi)^4}
\frac{d^4 p}{(2 \pi)^4}
\,(2 \pi)^4 \delta^4(k+p_1-p_3-q)\,(2 \pi)^4 \delta^4(k+p_1-l)\nonumber\\
&& \times  (2 \pi)^4 \delta^4(k-p_2+p)\,
\Biggl\{\frac{k^2-M^2}{D(k^2)}\, \frac{q^2-M^2}{D(q^2)}
\,\frac{l^2-m^2}{d(l^2)}\, \frac{p^2-m^2}{d(p^2)}\nonumber\\
&& - \frac{k^2-M^2}{D(k^2)}\, \frac{q^2-M^2}{D(q^2)}\,\pi^2
\delta(l^2-m^2)\,\delta(p^2-m^2)- \frac{M\,\Gamma}{D(k^2)}\,
\frac{M\,\Gamma}{D(q^2)} \,\frac{l^2-m^2}{d(l^2)}\,
\frac{p^2-m^2}{d(p^2)}\nonumber\\
&& - \frac{M\,\Gamma}{D(k^2)}\, \frac{q^2-M^2}{D(q^2)}
\,\frac{p^2-m^2}{d(p^2)}\,\pi\,\delta(l^2-m^2) -
\frac{M\,\Gamma}{D(k^2)}\, \frac{q^2-M^2}{D(q^2)}
\,\frac{l^2-m^2}{d(l^2)}\,\pi\,\delta(p^2-m^2)\nonumber\\
&& - \frac{k^2-M^2}{D(k^2)}\, \frac{M\,\Gamma}{D(q^2)}
\,\frac{p^2-m^2}{d(p^2)}\,\pi\,\delta(l^2-m^2) -
\frac{k^2-M^2}{D(k^2)}\, \frac{M\,\Gamma}{D(q^2)}
\,\frac{l^2-m^2}{d(l^2)}\,\pi\,\delta(p^2-m^2)\nonumber\\
&& + \frac{M\,\Gamma}{D(k^2)}\, \frac{M\,\Gamma}{D(q^2)}\,\pi\,
\delta(l^2-m^2)\,\pi\, \delta(p^2-m^2) \Biggr\}\,. \label{I4I}
\end{eqnarray}
    To rewrite the expression of Eq.~(\ref{I4I}) in a more convenient form
we consider the following integral,
\begin{eqnarray}
0 & = & i\int \frac{d^4 k}{(2 \pi)^4}\frac{d^4 q}{(2 \pi)^4}\frac{d^4 l}{(2 \pi)^4}
\frac{d^4 p}{(2 \pi)^4}
\,(2 \pi)^4 \delta^4(k+p_1-p_3-q)\,(2 \pi)^4 \delta^4(k+p_1-l)\nonumber\\
&& \times  (2 \pi)^4 \delta^4(k-p_2+p) \Delta_A'(k)\Delta_A'(q)\Delta_A(l)\Delta_R(p)\nonumber\\
& = & i \int \frac{d^4 k}{(2 \pi)^4}
\,\Delta_A'(k)\Delta_A'(p_1-p_3+k)\Delta_A(p_1+k)\Delta_R(p_2-k)\,,
\label{LoopI4zero}
\end{eqnarray}
take its imaginary part and obtain
\begin{eqnarray}
0 & = & \int \frac{d^4 k}{(2 \pi)^4}\frac{d^4 q}{(2 \pi)^4}\frac{d^4 l}{(2 \pi)^4}
\frac{d^4 p}{(2 \pi)^4}
\,(2 \pi)^4 \delta^4(k+p_1-p_3-q)\,(2 \pi)^4 \delta^4(k+p_1-l)\nonumber\\
& &\times  (2 \pi)^4 \delta^4(k-p_2+p) \,
\Biggl\{\frac{k^2-M^2}{D(k^2)}\, \frac{q^2-M^2}{D(q^2)}
\,\frac{l^2-m^2}{d(l^2)}\,
\frac{p^2-m^2}{d(p^2)}\nonumber\\
&& + \frac{k^2-M^2}{D(k^2)}\, \frac{q^2-M^2}{D(q^2)}\,\pi^2
\delta(l^2-m^2)\,\delta(p^2-m^2)\,\sigma_l\,\sigma_p-
\frac{M\,\Gamma\,\frac{k_0}{{\cal E}_k}}{D(k^2)}\,
\frac{M\,\Gamma\,\frac{q_0}{{\cal E}_q}}{D(q^2)}
\,\frac{l^2-m^2}{d(l^2)}\,
\frac{p^2-m^2}{d(p^2)}\nonumber\\
&& -  \frac{M\,\Gamma\,\frac{k_0}{{\cal E}_k}}{D(k^2)}\,
\frac{q^2-M^2}{D(q^2)}
\,\frac{p^2-m^2}{d(p^2)}\,\pi\,\delta(l^2-m^2)\,\sigma_l+
\frac{M\,\Gamma\,\frac{k_0}{{\cal E}_k}}{D(k^2)}\,
\frac{q^2-M^2}{D(q^2)}
\,\frac{l^2-m^2}{d(l^2)}\,\pi\,\delta(p^2-m^2)\,\sigma_p\nonumber\\
&& -  \frac{k^2-M^2}{D(k^2)}\, \frac{M\,\Gamma\,\frac{q_0}{{\cal
E}_q}}{D(q^2)}
\,\frac{p^2-m^2}{d(p^2)}\,\pi\,\delta(l^2-m^2)\,\sigma_l+
\frac{k^2-M^2}{D(k^2)}\, \frac{M\,\Gamma\,\frac{q_0}{{\cal
E}_q}}{D(q^2)}
\,\frac{l^2-m^2}{d(l^2)}\,\pi\,\delta(p^2-m^2)\,\sigma_p\nonumber\\
&& -  \frac{M\,\Gamma\,\frac{k_0}{{\cal E}_k}}{D(k^2)}\,
\frac{M\,\Gamma\,\frac{q_0}{{\cal E}_q}}{D(q^2)}\,\pi\,
\delta(l^2-m^2)\,\pi\, \delta(p^2-m^2)\,\sigma_l\,\sigma_p+{\cal
O}(\Gamma^2) \Biggr\}\,. \label{I4Izeroexp}
\end{eqnarray}
    Subtracting Eq.~(\ref{I4Izeroexp}) from Eq.~(\ref{I4I}), we obtain
\begin{eqnarray}
{\rm Im }[I_3] & = & \int \frac{d^4 k}{(2 \pi)^4}\frac{d^4 q}{(2 \pi)^4}\frac{d^4 l}{(2 \pi)^4}
\frac{d^4 p}{(2 \pi)^4}
\,(2 \pi)^4 \delta^4(k+p_1-p_3-q)\,(2 \pi)^4 \delta^4(k+p_1-l)\nonumber\\
&& \times  (2 \pi)^4 \delta^4(k-p_2+p) \, \Biggl\{
 - \frac{k^2-M^2}{D(k^2)}\,
\frac{q^2-M^2}{D(q^2)}\,\pi^2
\delta(l^2-m^2)\,\delta(p^2-m^2)(1+\sigma_l\,\sigma_p)\nonumber\\
&& - \frac{M\,\Gamma\,\left(1-\frac{k_0}{{\cal
E}_k}\,\frac{q_0}{{\cal E}_q}\right)}{D(k^2)}\,
\frac{M\,\Gamma}{D(q^2)} \,\frac{l^2-m^2}{d(l^2)}\,
\frac{p^2-m^2}{d(p^2)}\nonumber\\
&& - \frac{M\,\Gamma\,\left(1-\frac{k_0}{{\cal
E}_k}\,\sigma_l\right)}{D(k^2)}\, \frac{q^2-M^2}{D(q^2)}
\,\frac{p^2-m^2}{d(p^2)}\,\pi\,\delta(l^2-m^2)\nonumber\\
&& -  \frac{M\,\Gamma\,\left(1+\frac{k_0}{{\cal
E}_k}\,\sigma_p\right)}{D(k^2)}\, \frac{q^2-M^2}{D(q^2)}
\,\frac{l^2-m^2}{d(l^2)}\,\pi\,\delta(p^2-m^2)\nonumber\\
&& - \frac{k^2-M^2}{D(k^2)}\,
\frac{M\,\Gamma\,\left(1-\frac{q_0}{{\cal
E}_q}\,\sigma_l\right)}{D(q^2)}
\,\frac{p^2-m^2}{d(p^2)}\,\pi\,\delta(l^2-m^2)\nonumber\\
&& - \frac{k^2-M^2}{D(k^2)}\,
\frac{M\,\Gamma\,\left(1+\frac{q_0}{{\cal
E}_q}\,\sigma_p\right)}{D(q^2)}
\,\frac{l^2-m^2}{d(l^2)}\,\pi\,\delta(p^2-m^2)
+ \, {\cal O}(\Gamma^2) \Biggr\}\,. \label{I4Ifinal}
\end{eqnarray}
   Note that we have kept the second line in the brackets for completeness although
it is of ${\cal O}(\Gamma^2)$.
   In analogy to the step from Eq.~(\ref{a}) to Eq.~(\ref{LoopIImFinal}), one can
replace expressions like $\frac{q_0}{{\cal E}_q}$ by the corresponding functions $\sigma_q$ in
Eqs.~(\ref{LoopI3IFinal}) and (\ref{I4Ifinal}).
   Finally, the case $\Gamma =0$ is obtained by first replacing
$M\,\Gamma\to M\,\Gamma+\epsilon$ in all integrals above.
   Setting $\Gamma=0$ and then taking the limit $\epsilon\to0^+$, we exactly reproduce the
standard cutting formulas for loop integrals with real masses \cite{Cutkosky:1960sp,Peskin:1995ev}.


\end{document}